\documentclass[preprint,5p,times,twocolumn]{elsarticle}
\usepackage[utf8]{inputenc}
\usepackage{graphicx}
\usepackage{booktabs}
\usepackage{url}
\usepackage{lineno}
\usepackage{hyperref}

\journal{Nuclear Instruments and Methods in Research A}
\begin{document}
\begin{frontmatter}
\author[JLab]{J.~D.~Maxwell}\corref{cor1}
\ead{jmaxwell@jlab.org}
\author[MIT]{R.~G.~Milner}
\address[JLab]{Thomas Jefferson National Accelerator Facility, Newport News, VA}
\address[MIT]{Laboratory for Nuclear Science, MIT, Cambridge, MA}
\cortext[cor1]{Corresponding author}

\title{A Concept for Polarized $^3$He Targets for High Luminosity Scattering Experiments in \\High Magnetic Field Environments}

\begin{abstract}
We present the conceptual design of a polarized $^3$He target to be used for high luminosity scattering experiments within high magnetic field environments. This two-cell target will take advantage of advancements in optical pumping techniques at high magnetic field to create 60\% longitudinally polarized $^3$He gas in a pumping cell within a uniform magnetic field above 1\,T. By transferring the polarized gas to cryogenic target cell, the gas density is increased to create a target thickness suitable for high luminosity applications. We discuss the general design of this scheme, and plans for its application in Jefferson Lab's CLAS12 detector.

\end{abstract}

\end{frontmatter}

\section{Introduction}

Polarized Helium-3 offers an attractive target medium for accessing the neutron's spin properties. About 89\% of the time, $^3$He is in a spatially-symmetric S-state, where its two protons spins are anti-aligned in a spin singlet and the neutron spin is aligned with the nuclear spin. This makes polarized $^3$He an invaluable surrogate for a polarized free neutron target, and polarized $^3$He gas targets have been employed for scattering experiments in nuclear and particle physics~\cite{Gentile2017} at MIT, TRIUMF, IUCF~\cite{Lee1993}, SLAC, DESY~\cite{DeSchepper1998}, Mainz and Jefferson Lab. 

Two spin alignment techniques utilizing laser light have been used to create polarized $^3$He gas for such targets: metastability exchange optical pumping (MEOP) and spin-exchange optical pumping (SEOP). See~\cite{Gentile2017} for a recent, general review of these methods. In MEOP, an RF discharge excites a population of gas atoms into a metastable state which can be optically pumped to produce polarization, and this polarization is transferred to the larger ground state population by metastability-exchange collisions. In SEOP, rather than producing metastable atoms to pump, alkali metals which can be directly optically pumped are evaporated into the gas, and polarization is transferred via Fermi-contact hyperfine interactions between the alkali electron and the $^3$He nucleus. 

The largest operational difference between MEOP and SEOP is the gas pressure, and ultimately target density, at which these techniques can be performed. While MEOP has historically been limited to near 1\,mbar, SEOP is regularly used as high as 13 bar. This has made SEOP the technique of choice for high luminosity scattering experiments: SEOP targets were used for 13 experiments in Jefferson Lab's Hall A in the 6\,GeV era, and the first of 7 additional approved experiments has already run at 12\,GeV.

The prevalence of large magnetic fields in high energy and nuclear physics detector packages highlights a key limitation of current polarized $^3$He targets. High magnetic fields are used in spectrometers like prisms, separating particles of different momenta; this not only provides crucial information on the scattering interaction, but also constrains background particles from overwhelming detectors. Large acceptance spectrometers such as CLAS12 at JLab~\cite{Burkert}, sPHENIX at RHIC~\cite{CONNORS}, and CMS at the LHC~\cite{CMS} utilize strong solenoid fields around the interaction region as integral detector elements. Unfortunately, increasing wall relaxation makes SEOP less efficient at high magnetic fields~\cite{Chen_2011}, which has limited the availability of commonly used polarized $^3$He targets in such environments.

\section{Concept for a New Target}
We propose a novel, gaseous polarized $^3$He target based on {\bf high-field} MEOP techniques and double-cell cryogenic gas targets. In our proposed scheme, the atoms will be polarized within a high magnetic field and at room temperature before being transferred to a cold target cell, where the density of the gas is increased for scattering in the beam.  Our initial design has been aimed at producing polarized target nuclei inside the 5\,T solenoid of the CLAS12 detector in Jefferson Lab's Hall B, in support of a recently approved program of spin-dependent electron scattering from polarized $^3$He~\cite{maxwell2019, PAC48}.


\subsection{High-Field MEOP}

Since its invention by  Colegrove, Schearer and Walters in 1963~\cite{colegrove}, MEOP had typically been performed in holding fields near 30\,G. With increasing magnetic field, Zeeman splitting acts to decouple the electronic and nuclear spins, reducing the efficiency of MEOP and leading to a conventional wisdom that effective high-field MEOP was impossible above 0.1\,T. However, research at the Laboratoire Kastler Brossel at ENS in Paris, France---motivated by polarization for medical imaging in the presence of MRI magnets above 1\,T---showed that MEOP is not only possible at high magnetic fields, but these high fields allow high steady-state polarization at higher pressures~\cite{Courtade2000}. In 2004 they achieved 90\% steady state polarization at 1.5\,T and 1\,mbar, and by 2013 they had reached greater than 50\% polarization at 4.7\,T and 100\,mbar~\cite{nikiel2013}. While increasing the magnetic field does act to decouple the electron and nuclear spins, slowing transfer of polarization to the nucleus, this decoupling also inhibits polarization relaxation channels. In addition, the separation of hyperfine states creates highly absorbing lines, where clearer discrimination of polarizing transitions is possible with the pumping laser while avoiding depolarizing transitions.

\begin{figure}[h]
\centerline{\includegraphics[width = 0.9\columnwidth]{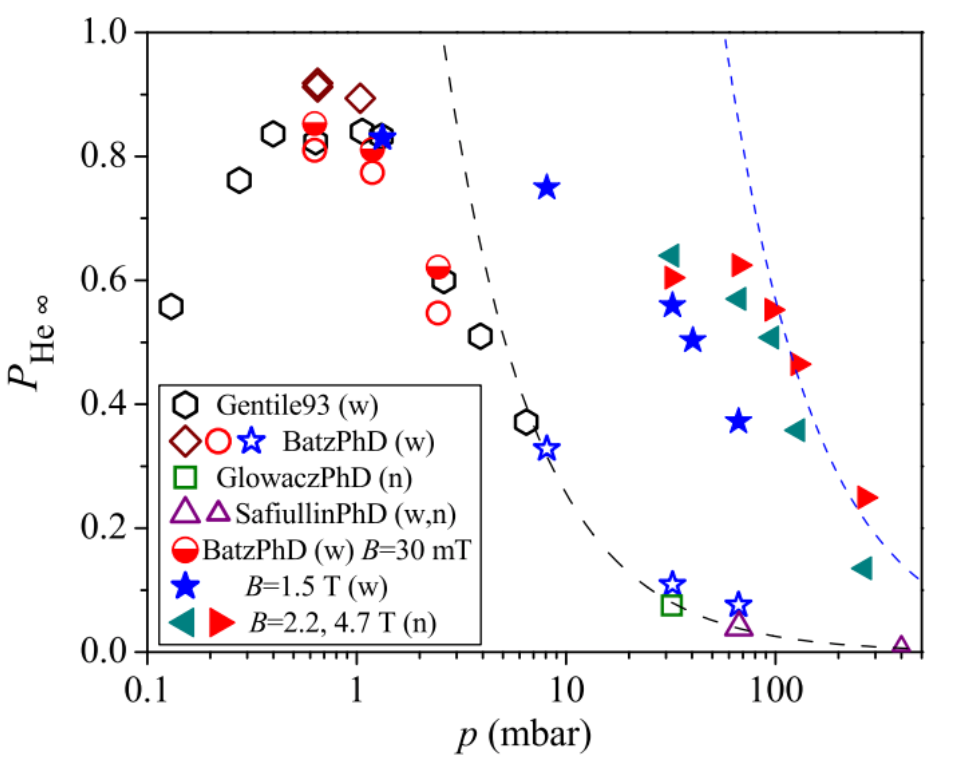}}
\caption{Variation with pressure of highest steady-state polarizations achieved by various groups at low fields (open symbols, 1 to 3 mT) and high fields (filled symbols, see the legend) from~\cite{Gentile2017}.}
\label{fig:ENS}
\end{figure}

The ENS group has studied the improvement of MEOP efficiency that comes with high magnetic field as pressures increase from 1 to 300\,mbar. Figure \ref{fig:ENS} shows their achieved steady-state polarization versus gas pressure for magnetic fields up to 4.7\,T. With no further improvement in the technique, polarizations as high as 60\% are possible at 100\,mbar, two orders of magnitude higher than typical MEOP pressures.
High-field MEOP techniques are already being applied for nuclear physics applications as the basis of a polarized $^3$He ion source for use at the Electron-Ion Collider~\cite{Max2016}, as pursued by the BNL Collider-Accelerator Department and these authors.


\subsection{Double-Cell $^3$He Target System}
\begin{figure}[h!]
\centerline{\includegraphics[width=1\columnwidth]{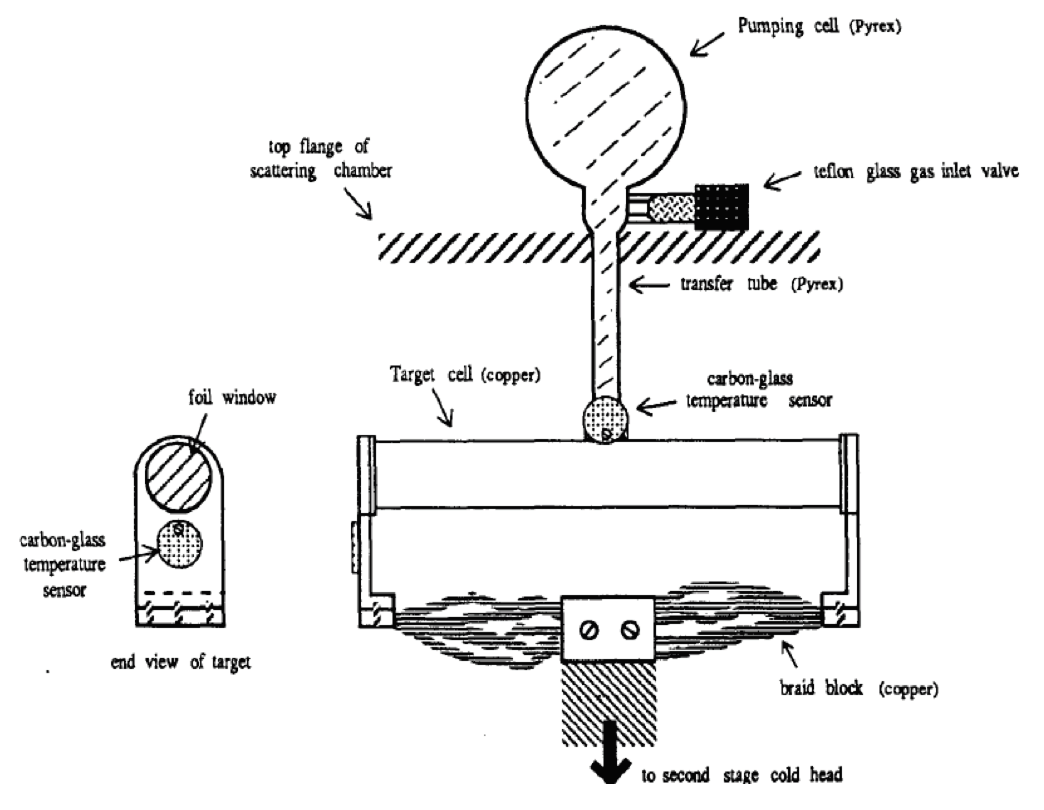}}
\caption{Schematic layout of the polarized $^3$He target double-cell system developed at Caltech~\cite{Milner1989} for Bates experiment 88-02.}
\label{fig:CIT-targ}
\end{figure}

A double-cell polarized $^3$He target was developed~\cite{Milner1989, jones1992measurement} at Caltech for Bates experiment 88-02 and was the first such target used for electron scattering experiments, including spin-dependent inclusive scattering in the quasielastic region~\cite{Wood1992}. It used an LNA-based, custom-built laser system which has been superseded by modern, commercially available, turn-key fiber-based systems.  The copper target cell had a circular cross section of diameter 2.54 cm and a length of 16 cm.   The interior of the target cell was coated with a thin layer of frozen nitrogen to reduce depolarization from interactions with the cell walls. The end windows were 4.6 $\mu$m thick copper foils which were epoxied to the target cell.  Fig.~\ref{fig:CIT-targ} shows a schematic layout of the Bates 88-02 target. 

During data taking, a total integrated charge of 1478 $\mu$A-hours was accumulated on this target. It was subsequently used in a second set of measurements~\cite{Gao1994,Hansen1995} of inclusive quasielastic spin-dependent electron scattering from polarized $^3$He at Bates in 1993.

\subsection{Proposed High-Field, Double-Cell Polarized Target}
By bringing high-field MEOP techniques to the traditional double-cell cryogenic polarized gas target design, a vast improvement in target density and experimental figure of merit can be realized. Where the Bates target operated at 40\% polarization near 2\,mbar with a 13\,K target cell, a new target operating in a high magnetic field could achieve 60\% polarization at 100\,mbar in a 5\,K target cell. While the gas density of such a target, at 5.4\,amg, would not surpass the 9\,amg typical of SEOP targets used at low fields with room temperature target cells, the ability to operate within high magnetic fields would enable new applications of polarized $^3$He targets.

Like the Bates 88-02 target, this new design would include two cells, now both held within a magnetic field above 1\,T: one pumping cell where the gas is polarized at room temperature, and one target cell through which the experimental beam may pass. The pumping cell is typical of most MEOP setups: glass with optically clearly ends to allow the passage of the pumping laser light, and coupled to electrodes to induce the RF discharge. Metastablility-exchange collision cross sections drop rapidly with decreasing temperature~\cite{Gentile2017}, so the pumping cell must be held near room temperature by heating elements. 


To further elucidate the implementation of a high-field MEOP target system, and to address specific challenges in its realization, we will provide an overview of the preliminary design of our first intended application of this idea, a polarized $^3$He target for Jefferson Lab's Hall B.

\section{Polarized $^3$He Target for CLAS12}

The CLAS12 spectrometer is a powerful tool designed to access the complete electromagnetic response for nucleons and nuclei in a wide range of kinematics~\cite{CLAS12TDR} and is aimed at the study of the internal dynamics and 3D imaging of the nucleon and quark hadronization processes. It is unique among JLab's standard detector equipment in providing a very large acceptance, while still allowing high electron luminosity up to $10^{35}$ cm$^{-2}$s$^{-1}$. CLAS12 offers excellent particle identification, and can detect neutrons and tagged particles. Its central detector includes a 5\,T superconducting solenoid to provide a magnetic field for the analysis of particle trajectories, and was designed to accommodate dynamically polarized solid targets such as NH$_3$ for protons and ND$_3$ for deuterons. CLAS12 is particularly well suited to the study of exclusive processes for Generalized Parton Distributions (GPDs) and semi-inclusive processes for Transverse Momentum Distributions (TMDs), where access to both current and target fragmentation regions---provided by a large acceptance---is critical. 

Despite the physics opportunities that a polarized $^3$He target would offer, one has never been used in Hall B in JLab's decades of experimentation.  While increasing the holding field above the typical 100\,G for SEOP pumping has shown some benefit---the alkali-alkali relaxation rates have been shown to reduce by a factor of 2 near 500\,G~\cite{kadlecek}---increases in wall relaxation with increasing field outstrip this effect~\cite{Chen_2011} to make high-field SEOP untenable. Using a  conventional, low-field SEOP target in CLAS12 would require either removing the central solenoid, moving the target upstream of the spectrometer or transferring gas polarized at low field into the solenoid's 5\,T field. Removing the solenoid or moving the interaction region out of CLAS12 would require major modification to the configuration and abilities of the spectrometer, as well as the analysis tools used to extract data. We have studied the transfer of gas through depolarizing field gradients~\cite{maxwell2014}, and it can be challenging to accomplish without significant loss of polarization. High-field MEOP offers a clear path to alleviate these constraints without significant changes to the detector's configuration. 

\begin{figure*}[t]
\centerline{\includegraphics[width=1.6\columnwidth]{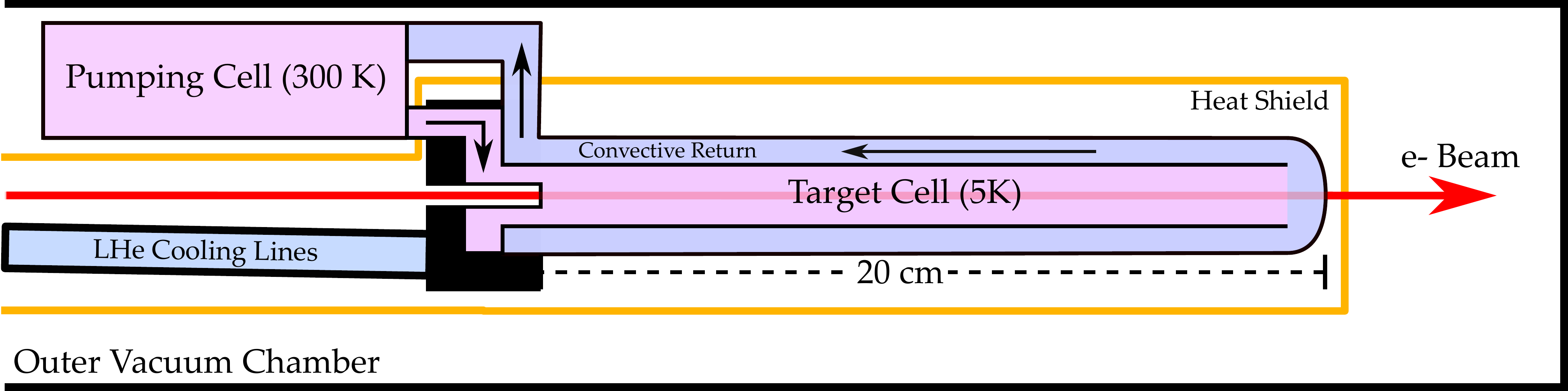}}
\caption{Schematic layout of a two cell polarized $^3$He target system in CLAS12.}
\label{fig:Sideview}
\end{figure*}

Our design for a high-field polarized $^3$He target would follow the two-cell, cryogenic concept which we have outlined, while adapting to the constraints and abilities of the CLAS12 spectrometer. Both cells of the system must fit within the available 10\,cm cylindrical space while allowing the passage of the beam through the center. 
Figure~\ref{fig:Sideview} shows a diagram of the proposed design, with two gas cell volumes in convective contact---one cooled by a liquid helium heat exchanger, the other heated and optically pumped. Using 100\,mbar gas in a 20\,cm long aluminum target cell at 5\,K will result in a target thickness of $2.9\times10^{21}$ $^3$He/cm$^2$, which at a beam current of 0.5 $\mu A$ will produce a per nucleon luminosity of  $2.7\times10^{34}$ nucleons/cm$^2$/s. Beam windows on the entrance and exit of the cell will be 25\,$\mu$m thick aluminium foils, and an additional beam window on the outer vacuum chamber will be 50\,$\mu$m thick.
Table~\ref{tab:Comp} compares the achieved specification for the Bates 88-02 target with those proposed for the CLAS12 target based on demonstrated high-field optical pumping performance. 

\begin{table}[h]
\begin{center}
\begin{tabular}{lcc}
\toprule
 &  Bates 88-02   &  CLAS12 \\
 Parameter         &   Target   &  Target  \\
          & Achieved & Proposed\\
          \midrule
Pumping cell pressure (mbar) & 2.2 & 100\\
Pumping cell volume (cm$^3$) & 200& 200\\
Target cell volume (cm$^3$) & 79& 16\\
Target cell length (cm) & 16& 20\\
Atoms in pumping cell  &$4.6 \times 10^{18}$ & $5.2 \times 10^{20}$\\
Atoms in target cell  & $9.6 \times 10^{19}$ & $2.3 \times 10^{21}$\\
Holding field (T)  &0.003 & 5\\
Polarization & 38\% & 60\% \\
Electron beam energy (GeV) & 0.574 & 10\\
Cell temperature (K) & 13 & 5\\
Target thickness ($^3$He/cm$^2$) &$2 \times 10^{19}$ & $3 \times 10^{21}$\\
Beam current ($\mu$A) & 10 & 0.5\\
Luminosity on He ($^3$He/cm$^2$/s) & $3 \times 10^{33}$& $9 \times 10^{33}$\\
\bottomrule
\end{tabular}
\end{center}
\caption{Comparison of specifications for the Bates 88-02 target~\cite{Gao1994} and the proposed CLAS12 polarized $^3$He target.}
\label{tab:Comp}
\end{table}


\subsection{Cryogenics and Heat Load}

Modern pulse-tube cryo-cooler systems provide sufficient power to allowing cooling of the cryogenic target cell and overcome heating from the electron beam. We plan to use a Cryomech PT-420 pulse tube, which can provide 2\,W of cooling power at 4.2\,K, to condense liquid helium to circulate through a heat exchanger on the target cell to maintain the temperature below 5\,K. 

Heat loads from various sources must be considered to ensure it stays below the 2\,W which can be delivered by the pulse-tube.
The power deposition due to energy loss by the beam in the target gas itself will be roughly 100\,mW, while the foil windows are estimated contribute an additional 7 mW. The glass transfer tube between the cells will also conduct heat to the target cell, creating another roughly 30\,mW of heat to remove. 

As MEOP does not operate effectively at low temperature, the pumping cell will be kept at room temperature. A simple calculation of heat exchange via conduction in a static gas gives roughly 35\,mW of heat. To look for an upper limit on the heat load from gas transfer, we can assume that all the gas in the pumping cell is cooled to 5\,K every 8 seconds (an estimated transfer time from the Bates target and overestimate at these pressures), which would result in roughly 350 mW of heat. 

The transfer line between the cells will need to be optimized to balance the heat load from gas diffusion and the supply of polarized gas to the target cell, perhaps using the temperature differential between cells to create convective flow through two transfer lines. All together, these heat loads are within the 2\,W of cooling power of the system at 4.2\,K, however a detailed heat analysis will be performed before the design is finalized.

\subsection{Depolarization Effects}

The main sources of polarization relaxation come from wall interactions, transverse magnetic field gradients, and ionization in the beam. To avoid depolarization on the cell walls, the room temperature pumping cell will be made of borosilicate glass, and the aluminum target cell will be coated with a cryogenic layer of H$_2$, which has been shown to yield days long relaxation times between 2 and 6\,K~\cite{lefevre}. The transfer line itself will be glass transitioning to metal, where all metal parts will be cold enough to facilitate the cryogenic coating.

\begin{figure}[h!]
\centerline{\includegraphics[width=\columnwidth]{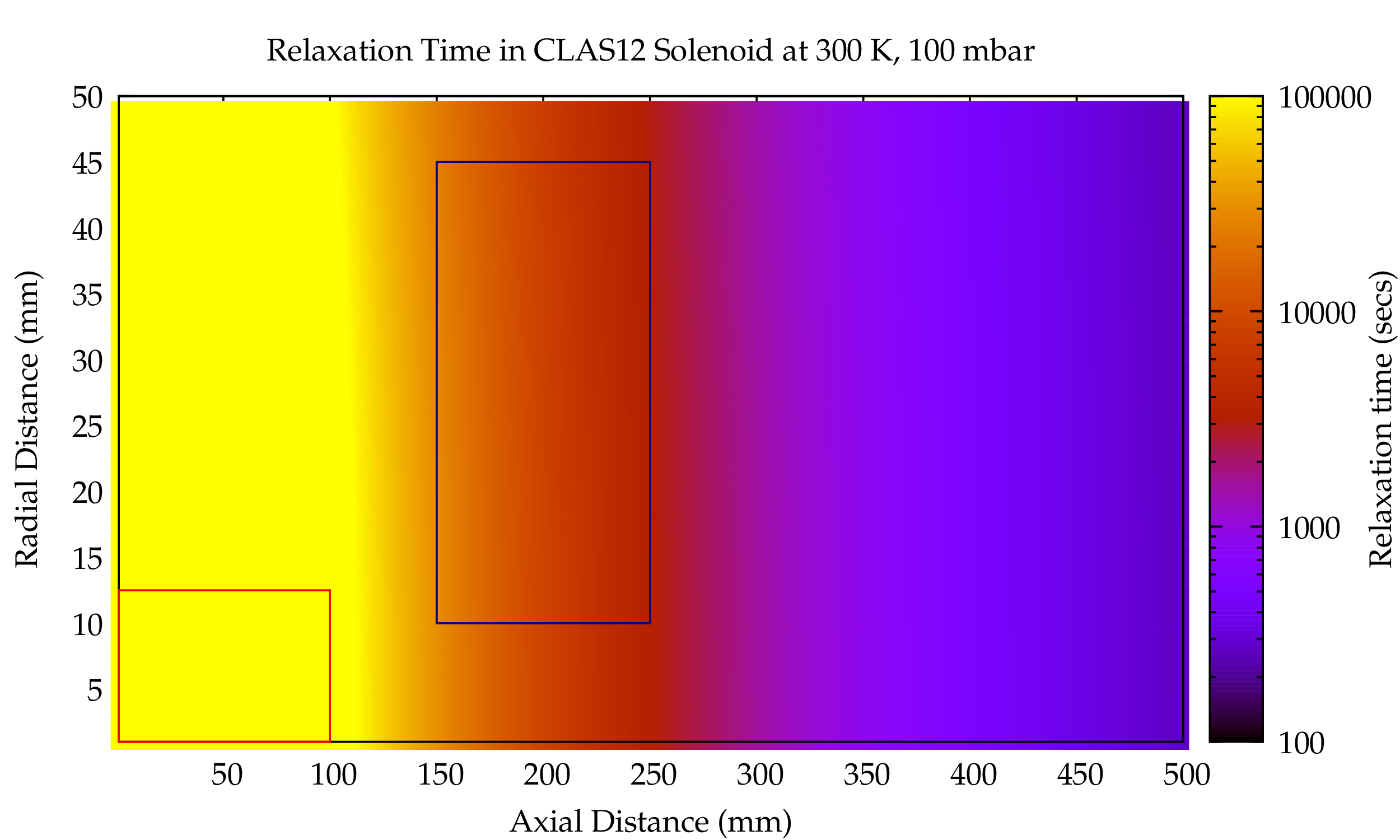}}
\caption{Preliminary map of $^3$He relaxation time due to transverse field gradients in the CLAS12 solenoid, showing distance from the center of the solenoid. This map assumes gas at 100\,mbar and 300\,K, and shows candidate locations for a pumping cell (blue box) and target cell (red box).}
\label{fig:map}
\end{figure}

The relaxation time $T_g$ due to transverse gradients in a holding field directed along the $z$-direction is given by
$$
\frac{1}{T_g} = \frac{\langle v^2 \rangle }{3}\frac{|\nabla B_x|^2 +|\nabla B_y|^2}{B^2_0} \left(\frac{\tau_c}{1+\omega_0^2 \tau_c^2} \right)
$$
where $B_0$ is the holding field, $\tau_c$ is the meantime between atomic collisions and $\omega_0 = \gamma B_0$ is the Larmor frequency for the magnetic field~\cite{schearer}.  For $^3$He the gyromagnetic ratio, $\gamma$, is 3.24 kHz/G.  The mean collision rate has been measured as a function of pressure at 300 K and determined as
$$
\tau_c = (2.2 \pm 0.2) \times 10^{-7} p^{-1} \ {\rm sec} ,
$$
where $p$ is the pressure in Torr.  $\langle v^2 \rangle = 3kT/m$ is the mean square thermal velocity of the atoms.
The relaxation rate from magnetic field gradients decreases with the temperature because the atoms move more slowly, experiencing smaller fluctuations in the field in a given amount of time. In a double-cell system, where the target cell is cooled and the pumping cell is operated at room temperature, the effect of the field gradients is more important for the warmer pumping cell. Figure \ref{fig:map} shows a map of relaxation time in the central axial and radial space of CLAS12 for 300\,K and 100\,mbar $^3$He gas, showing candidate locations for pumping and target cells. For the 5\,K target cell, the gas particle velocity will be much lower, making relaxation times much higher than shown in this map. We expect that depolarization due to field gradients will be negligible if the pumping cell is located inside the solenoid.

\begin{figure}[h!]
	\includegraphics[width=\columnwidth]{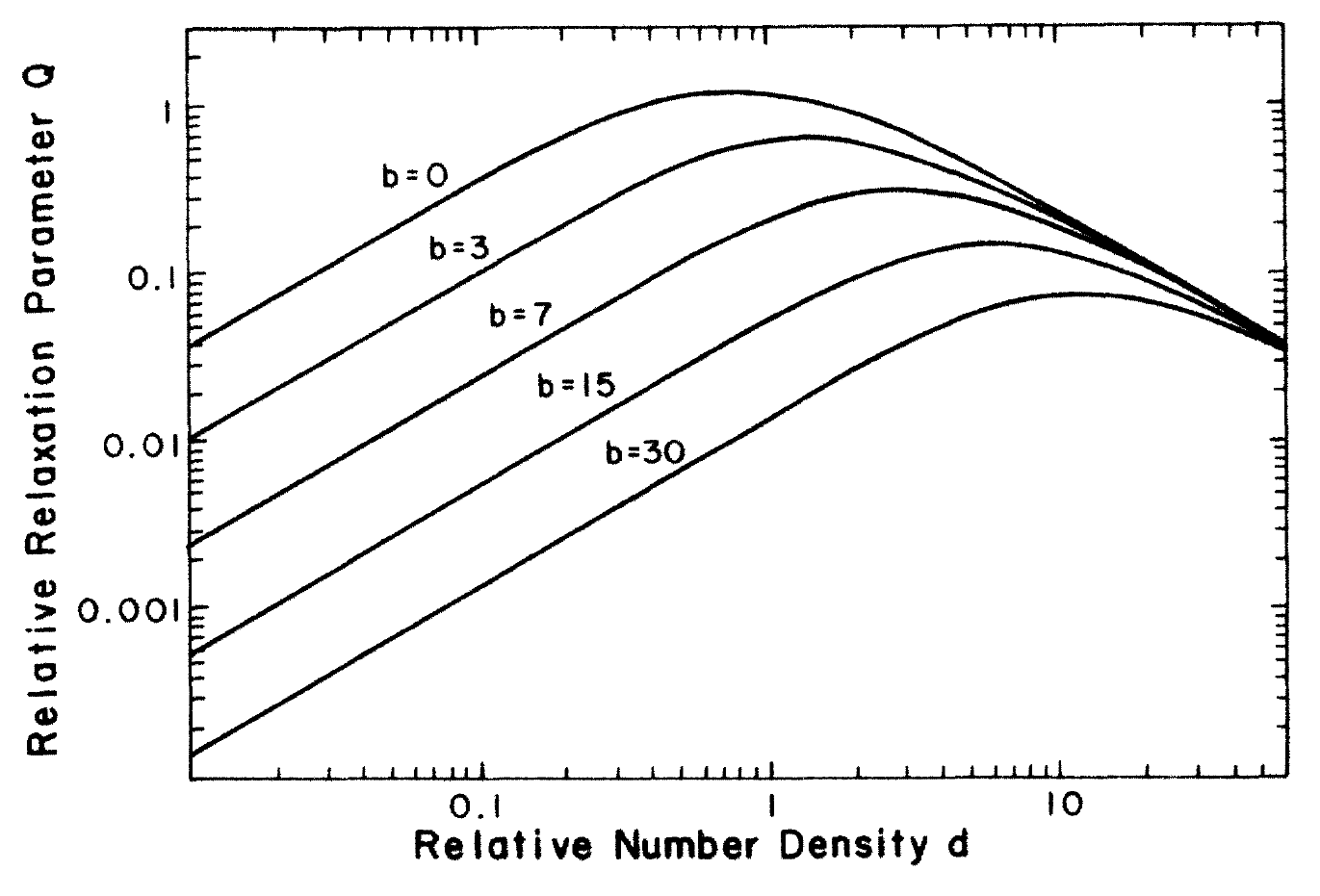}
\caption{the relative relaxation rate vs. number density, showing the decrease in the relaxation rate with increased magnetic field~\cite{bonin1988relaxation}. }
\label{fig:bonin}
\end{figure}

Ionizing radiation can induce spin relaxation through the production of molecular $^3$He$_2^+$. This effect was studied extensively for the Bates 88-02 target, and was found to create a 2000 second relaxation time in 2.6\,mbar gas under a beam current of 5$\mu$A. While the molecular production increases with density, increasing the magnetic field reduces the depolarization rate from to diatomic molecules. Figure \ref{fig:bonin} gives the relative relaxation rate vs.~gas number density, showing the strong effect of increased magnetic field, here expressed as a relative value $b$, the ratio of the holding field over the atom's characteristic field. An increase in field from 10\,G to 200\,G, reduces the relaxation rate by two orders of magnitude as the rotational angular momentum spin is decoupled from the total molecular-ion spin~\cite{bonin1988relaxation}.  
The rate of communication between the cells, delivering polarization from the pumping cell to the target cell, will be studied extensively to ensure that the design promotes sufficient convection to maintain high polarization in the target cell.




\subsection{Polarization Measurement}

\begin{figure}[!h]
\centerline{\includegraphics[width=\columnwidth]{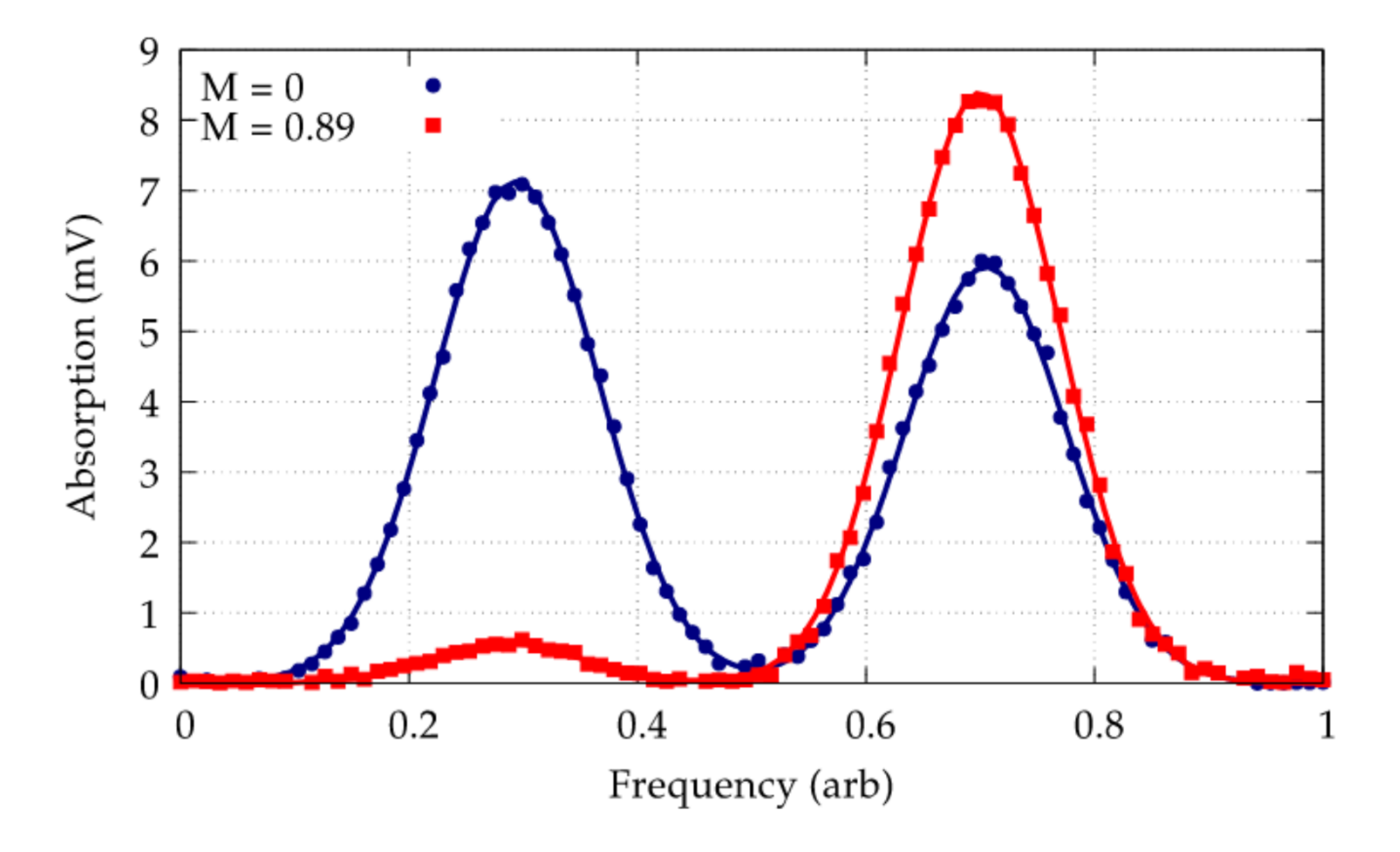}}
\caption{Absorption signals for populations $a_1$ (left) and $a_2$ (right) measured  with ground state nuclear polarization at 0 and 89\%, using a 1 torr sealed cell at 3 T~\cite{maxwell2020}.}
\label{fig:NIMpol}
\end{figure}

At low field, gas polarization during MEOP can be easily measured by observing the circular polarization of 667\,nm light emitted in the discharge~\cite{maxwell2014-2}, however, the decoupling of electronic and nuclear spins at high field reduces the accuracy of such measurements at high field. At high magnetic field, the Zeeman separation of the states can be utilized via the absorption of a secondary probe laser as it passes through the discharge. The population of two particular 2$^3$S sub-levels can be monitored by sweeping the probe laser frequency, providing a measure of the ground state polarization~\cite{Gentile2017, nikiel2013}. These sublevels are chosen to avoid the states under active optical pumping.  At spin-temperature equilibrium, the populations of these probed states, here $a_1$ and $a_2$, will satisfy $a_2/a_1 = (1+M)/(1-M)$.  An absolute measure of the nuclear polarization $M$ of the ground states can be formed from the change in the ratio $r=a_2/a_1$ of the absorption signal amplitudes for these sublevels during MEOP, as calibrated by their ratio $r_0$ when not polarized ($M=$ 0):
\begin{equation}
M = \frac{r/r_0-1}{r/r_0+1} \ .
\end{equation}
Because only ratios of spectral amplitudes are involved, all experimental parameters affecting the absolute signal intensities are canceled out, making this a robust measurement.  Figure~\ref{fig:NIMpol} shows sample absorption signals for a high value of polarization~\cite{maxwell2020}.

Probe laser polarimetry will be effective in the pumping cell, where metastable atoms will absorb the probe laser light, but this techniques is not possible in the aluminum target cell. While pulse NMR methods may be pursued to monitor the polarization of the gas as it travels to and returns from the target cell, we intend to take advantage of a method developed for the Bates 88-02 target to infer the polarization in the target cell based on measurements of the polarization in the pumping cell. By solving rate equations describing the rate of polarization relaxation in each cell and the rate of diffusion between the cells, the target cell polarization can be inferred using measurements of polarization and relaxation in the pumping cell over time~\cite{ Milner1989, jones1992measurement}.

\subsection{Transversely Polarized Target}
We are investigating a concept to use this new target design to provide polarization transverse to the incident direction of the beam using superconducting shielding.
Taking advantage of the progress of fellow Jefferson Lab scientists from the HD-Ice group, we will adapt their plan to cancel the CLAS12 solenoidal field and produce a transverse holding field with bulk, superconducting MgB$_2$~\cite{Statera}. Our concept would polarize in CLAS12's 5\,T field at 100\,mbar, and transfer into the target cell, held within the MgB$_2$ shield. Polarized $^3$He requires only a small ($\sim50$G) holding field to maintain polarization, much less than the 1.5\,T planned for HD-Ice. In our scheme, the longitudinally polarized  $^3$He spins will rotate adiabatically in transit, following a rotating  field  trapped into the bulk superconductor, arriving transversely aligned to the beam at the target cell.  The value of transversely polarized physics in CLAS12 would be immense, and further research into target methods is needed to realize it.



\section*{Conclusion}
The combination of novel, high-field MEOP techniques and a cryogenic double-cell target offers a new approach to high-luminosity scattering experiments on polarized $^3$He nuclei for applications requiring a high magnetic field. A set of measurements of neutron spin structure utilizing this new target design within the CLAS12 detector in Hall B has been conditionally approved by the Jefferson Lab programming advisory committee to ~\cite{PAC48}. This experiment will explore the transverse momentum dependence of
the longitudinal spin structure of the neutron and investigate nuclear effects in SIDIS, comparing polarized $^3$He target scattering to previous measurements with polarized deuterium targets. Construction of a high-field MEOP apparatus for the development of this concept has begun at Jefferson Lab, with the aim of satisfying the programming committee's conditions for the full approval of a longitudinally polarized $^3$He target and exploring the feasibility of transverse polarization.

\section*{Acknowledgements}

We thank Pierre-Jean Nacher and Genevieve Tastevin from the Laboratoire Kastler Brossel, {\'E}cole Normale Sup{\'e}rieure, Paris, France for sharing their data and insights into high field MEOP at high pressures.  We thank Cyril Wiggins and Marco Battaglieri from Hall B, Jefferson Laboratory, and Victoria Lagerquist from ODU for providing us with important technical information. We gratefully acknowledge Chris Keith, Dave Meekins and James Brock from the Jefferson Laboratory Target Group for several helpful discussions. This material is based upon work supported by the U.S. Department of Energy, Office of Science, Office of Nuclear Physics under contracts  DE-AC05-06OR23177 and DE-FG02-94ER40818.

\bibliographystyle{elsarticle-num}
\bibliography{high_field_meop_target}

\end{document}